\documentclass[12pt]{article}

\usepackage{amssymb,amsmath}
\usepackage{graphics}
\usepackage{epsfig}
\usepackage{amsfonts}
\usepackage{mathrsfs}
\usepackage{psfrag}

\usepackage{amscd}

\usepackage{mathrsfs}
\usepackage{cite}
\begin{document}
\begin{center}
{\Large Duality covariant variables for STU-model in presence of non-holomorphic corrections}
\end{center}
\centerline{\large \rm Shamik Banerjee, Rajesh Kumar Gupta}

\centerline{\large \it Harish-Chandra Research Institute}

\centerline{\large \it  Chhatnag Road, Jhusi,
Allahabad 211019, INDIA}
\vspace*{1.0ex}

\centerline{E-mail:
bshamik@mri.ernet.in, rajesh@mri.ernet.in}

\vspace*{8.0ex}
\begin{abstract}
It is known that non-holomorphic corrections are necessary in order to get duality invariant free energy and entropy function. However the present methods of incorporating non-holomorphic corrections are in conflict with special geometry properties of moduli space. The moduli fields do not transform in the duality covariant way and their duality transformation also involves the graviphoton field strength. In the present note we construct duality covariant moduli fields for STU-model perturbatively in powers of the graviphoton field strength and demonstrated their existence upto second order.
\end{abstract}
\vspace*{8.0ex}
\section{Introduction}
The four dimensional $N=2$ supergravity lagrangian \cite{0007195} coupled non-minimally to $(n+1)$-vector multiplets is described by a holomorphic prepotential $F(Y^I,\Upsilon)$, which is a homogeneous function of degree 2 in $Y^I(I=0,....,n)$ and $\Upsilon$. Here the $Y^I$ denotes the (rescaled) complex scalar field which sits in the abelian vector multiplet and $\Upsilon$ denotes the (rescaled) square of the auxiliary, antiselfdual, antisymmetric Lorentz tensor $T_{ab}^{ij}$ which sits in the weyl multiplet. The $\Upsilon$ dependent term in the prepotential gives rise to higher derivative curvature terms in the effective action. Under electric/magnetic duality transformation $Sp(2n+2,\mathbb{Z})$, $(Y^I,F_I(Y,\Upsilon))$ transform as a symplectic vector.\\Vector multiplet scalars assume fixed values at the black hole horizon determined by the attractor equation. The attractor equation relates two symplectic vectors $(p^I,q_I)$ and $(Y^I,F_I(Y,\Upsilon))$ and has the following form
\begin{equation}\label{attreqn}
Y^I-\bar {Y}^I=ip^I ,\quad F_I(Y,\Upsilon)-\bar{F}_I(\bar Y,\bar \Upsilon)=iq_I, \quad \Upsilon=-64
\end{equation}
 where $(p^I,q_I)$ denotes the magnetic and electric charges of the black hole. This equation is manifestly symplectic invariant.
The macroscopic entropy of a static, supersymmetric black hole computed \cite{0007195} from the effective lagrangian is given by
\begin{equation}\label{entropy}
S_{macro}(p,q)=\pi\left[{\vert Z\vert}^2 -256 Im(F_{\Upsilon}(Y,\Upsilon))\right]_{\Upsilon=-64}
\end{equation}
where
\begin{equation}
\vert Z \vert^2=p^IF_I(Y,\Upsilon)-q_IY^I
\end{equation}
There exist a proposal \cite{0601108} for the BPS entropy function whose stationary points are attractor equations and the value of the entropy function at the attractor points equals the macroscopic entropy (\ref{entropy}).
In terms of the entropy function, the macroscopic degeneracy of black hole is given by
\begin{equation}\label{degeneracy}
d(p,q)_{macro}=\int d(Y^I+\bar Y^I)d(F_I+\bar F_I)e^{\pi \Sigma(Y,\bar Y,p,q)}
\end{equation}
where the entropy function is given by
\begin{equation}\label{entropyfn}
\Sigma(Y,\bar Y, \Upsilon, \bar \Upsilon)=\mathscr{F}(Y,\bar Y,\Upsilon, \bar \Upsilon)-q_I(Y^I+\bar Y^I)+p^I(F_I+\bar F_I)
\end{equation}
Here $\mathscr{F}(Y,\bar Y,\Upsilon, \bar \Upsilon)$ denotes the free energy and is given by 
\begin{equation}\label{freeenrg}
\mathscr{F}(Y,\bar Y,\Upsilon, \bar \Upsilon)=-i\left(\bar Y^IF_I-Y^I\bar F_I\right)-2i\left(\Upsilon F_{\Upsilon}-\bar \Upsilon\bar F_{\Upsilon}\right)
\end{equation}
Since $(Y^I,F_I(Y,\Upsilon))$ transforms as a symplectic vector under symplectic transformation, the free energy (\ref{freeenrg}), the entropy function (\ref{entropyfn}) and hence the $d(p,q)_{macro}$ are symplectic invariant. \\A subgroup of the symplectic transformation is called ``duality symmetry'' when the function $F(Y,\Upsilon)$ remains unchanged i.e. $\tilde F(\tilde Y,\Upsilon)=F(\tilde Y,\Upsilon)$. This implies that the substitution of $Y^I\rightarrow\tilde Y^I$ into the derivatives $F_I(Y,\Upsilon)$ will induce the correct transformations on the $F_I(Y,\Upsilon)$. \\
The attractor equations (\ref{attreqn}) follow from the extremization of (\ref{entropyfn}) with respect to $Y^I$ keeping $\Upsilon$ fixed and the macroscopic entropy (\ref{entropy}) is given by
\begin{equation}
 S_{macro}(p,q)=\pi\Sigma\vert_{attractor}
\end{equation}
The entropy formula (\ref{entropy}) was based on the effective action approach. In this approach, coupling functions multiplying the higher derivative terms, like square of Riemann tensor, is proportional to Im$F_{\Upsilon}(Y,\Upsilon)$. The physical effective couplings in one P.I. effective action which one obtain by integrating out the fields at one loop are different from wilsonian coupling \cite{9906094,0601108} because of presence of infrared divergence due to massless fields. Although one does not know the complete effective action with non-local terms it is clear that the holomorphic function $F(Y,\Upsilon)$ which give rise to local higher derivative terms will not give duality invariant entropy (\ref{entropy}) and entropy function(\ref{entropyfn}). However one can still make the entropy function (\ref{entropyfn}), that appears in the expression (\ref{degeneracy}) for $d(p,q)$, duality invariant by modifying the function $F(Y,\Upsilon)$. One does this by relaxing the holomorphicity properties of the function $F(Y,\Upsilon)$ and taking into account a non-holomorphic corrections to $F(Y,\Upsilon)$.\\
The nonholomorphic coorections are added to the entropy function by replacing $F(Y,\Upsilon)$ by\cite{0601108,0808.2627}
\begin{equation} 
F(Y,\bar Y,\Upsilon,\bar \Upsilon)= F^{(0)}(Y)+2i\Omega(Y,\bar Y,\Upsilon,\bar\Upsilon)
\end{equation} 
 Here $\Omega(Y,\bar Y,\Upsilon,\bar\Upsilon)$ is a real, homogeneous function of degree 2 and $F^{(0)}(Y)$ is $\Upsilon$ independent part of the prepotential. The homogeneity condition implies that
\begin{equation}\label{homog}
Y^I\Omega_I+\bar Y^I\Omega_{\bar I}+2\Upsilon\Omega_\Upsilon+2\bar\Upsilon\Omega_{\bar\Upsilon}=2\Omega
\end{equation}
The form of $\Omega$ should be such that under duality transformation the substitution of $Y^I\rightarrow \tilde Y^I$ into the derivatives $F_I(Y,\bar Y,\Upsilon,\bar \Upsilon)$ will induce the correct duality transformation on $F_I(Y,\bar Y,\Upsilon,\bar \Upsilon)$.\\
Thus one obtains the duality invariant expressions by replacing $F(Y,\Upsilon)$ by $F(Y,\bar Y,\Upsilon,\bar \Upsilon)$ into (\ref{degeneracy}-\ref{freeenrg}). For example the duality invariant free energy (\ref{freeenrg}) becomes
\begin{equation}\label{modfrenrg}
\mathscr{F}(Y,\bar Y,\Upsilon, \bar \Upsilon)=-i(\bar Y^IF^{(0)}_I-Y^I\bar F^{(0)}_I)-2(Y^I-\bar Y^I)(\Omega_I-\Omega_{\bar I})+4\Omega
\end{equation}
Here we have used the eqn (\ref{homog}).\\
Recently the form of $\Omega$ has been determined for specific $N=2$-models \cite{0808.2627}. In \cite{0808.2627}, the form of $\Omega$ was guessed from the transformation rule for the derivatives of $\Omega$. These transformation rule was obtained by assuming that the duality transformation constitute an invariance of the model and duality transformation of the $Y^I$ induce the expected transformation of $F_I=F^{(0)}_I+2i\Omega_I$.\\
In presence of non-holomorphic corrections determined by $\Omega$, the new duality transformation mixes $Y^I$ and $\bar Y^I$. However this is in contrast with string theory where duality transformations act holomorphically on the physical moduli. Furthermore, string theory duality transformations of moduli fields does not involve $\Upsilon$ dependent terms. Hence the variables used in \cite{0808.2627} are not string theory variables. Furthemore, since the duality transformation mixes various powers of $\Upsilon$, the terms in the expression of free energy for a given power of $\Upsilon$ is not duality invariant.\\
In this note, we shall try to obtain new variables ${Y^{\prime}}^I$  for STU-model which ,even in presence of nonholomorphic corrections, transform as string theory variables. The old variables $Y^I$ are function of the new variables ${Y^{\prime}}^I$. These functions are detetrmined by requiring that under the standard duality transformation of ${Y^{\prime}}^I$, the $Y^I$ transform as in \cite{0808.2627}. These functions are obtained as a power series in $\Upsilon$. In the limit $\Upsilon\rightarrow 0$, ${Y^{\prime}}^I$ coincides with $Y^I$.
Furthermore the terms in the expression of free energy for a given power in $\Upsilon$ written in the new variables ${Y^{\prime}}^I$ are duality invariant. Although we will explicitly work in STU-model, the method can be applied for other models like FHSV-model also.\\
This paper is organised as follows. In section 2 we briefly introduce STU-model and determine the non-holomorphic corrections upto $\Upsilon^2$. We will closely follow \cite{0808.2627}. In section 3 we will introduce new variables which transform in the standard way under duality transformation. In section 4 we re-expressed the free energy in terms of these new variables. In section 5 we present our conclusion.
\section{STU-model and Nonholomorphic Corrections} 
The STU-model \cite{0711.1971,9508064} is based on four fields $Y^0,Y^1,Y^2$ and $Y^3$ of which later three appear symmetrically. The special coordinates S, T and U are defined as
\begin{equation}\label{stu}
S=\frac{-iY^1}{Y^0},\quad T=\frac{-iY^2}{Y^0},\quad U=\frac{-iY^3}{Y^0}.
\end{equation}
The tree level prepotential for the STU-model is given by
\begin{equation}\label{treeprepot}
F^{(0)}(Y)=-\frac{Y^1Y^2Y^3}{Y^0}=i(Y^0)^2STU.
\end{equation}
The complete duality group of the STU-model is
\begin{equation}\label{group}
(\Gamma(2)_S\otimes\Gamma(2)_T\otimes\Gamma(2)_U)\times \mathbb{Z}_2^{T-U}\times\mathbb{Z}_2^{S-T}\times\mathbb{Z}_2^{S-U}
\end{equation}
where $\Gamma(2)\subset SL(2;\mathbb Z)$ with $a,d\in2\mathbb Z+1$ and $b,c\in2\mathbb Z$, with $ad-bc=1$.\\

The action of S-duality group $\Gamma(2)_S$ is defined as
\begin{equation}\label{dualtransf}
\begin{split}
& Y^0\rightarrow dY^0+cY^1 , \qquad F^{(0)}_0\rightarrow aF^{(0)}_0-bF^{(0)}_1\\
& Y^1\rightarrow aY^1+bY^0 ,\qquad F^{(0)}_1\rightarrow dF^{(0)}_1-cF^{(0)}_0\\
& Y^2\rightarrow dY^2-cF^{(0)}_3 ,\qquad F^{(0)}_2\rightarrow aF^{(0)}_2-bY^3\\
& Y^3\rightarrow dY^3-cF^{(0)}_2 ,\qquad F^{(0)}_3\rightarrow aF^{(0)}_3-bY^2.
\end{split}
\end{equation}
One has similar expression for the T(U)-duality transformation by interchanging the labels $1\leftrightarrow 2 (1\leftrightarrow 3)$.\\
Next we want to include nonholomorphic corrections to STU-model. As mentioned above, this is done by considering the new prepotential
\begin{equation}\label{prepot}
F(Y,\bar Y,\Upsilon,\bar\Upsilon)=i(Y^{0})^2STU+2i\Omega(Y^{0},Y^1,Y^2,Y^3,\bar{Y^0},\bar{Y^1},\bar{Y^2},\bar{Y^3},\Upsilon,\bar\Upsilon).
\end{equation}
Here $\Omega$ is a real homogeneous function of degree 2 in $Y$ and $\Upsilon$ and contains the non-classical contribution. Throughout our analysis we will assume that $\Upsilon$ is real - this can be achieved by working in an appropriate gauge condition.\\
Assuming that $\Omega$ is analytic at $\Upsilon=0$, one can expand it as 
\begin{equation}\label{omegaexp}
\Omega(Y^{0},S,T,U,\bar{Y^0},\bar{S},\bar{T},\bar{U},\Upsilon)=\Upsilon \Omega^{(1)}+\displaystyle\sum_{g=2}^{\infty} \Upsilon^g \Omega^{(g)}
\end{equation}
Since $\Omega$ is a homogeneous functon of degree 2 , $\Omega^{(1)}$ will depend on $S,T,U$ and their complex conjugate but will not depend on $Y^0$ and $\Omega^{(g)}$ will be a homogeneous function of degree $2-2g$ in the $Y^{I}$'s.\\
The holomorphic derivatives of function $F$ with respect to $Y^I$ are
\begin{equation}
\begin{split}
&F_0=-iY^0STU-\frac{2i}{Y^0}\left[-Y^0\frac{\partial}{\partial Y^0}+S\frac{\partial}{\partial S}+T\frac{\partial}{\partial T}+U\frac{\partial}{\partial U}\right]\Omega\\
&F_1=Y^0TU+\frac{2}{Y^0}\frac{\partial\Omega}{\partial S}\\
&F_2=Y^0SU+\frac{2}{Y^0}\frac{\partial\Omega}{\partial T}\\
&F_3=Y^0ST+\frac{2}{Y^0}\frac{\partial\Omega}{\partial U}.\\
\end{split}
\end{equation}
Under $\Gamma(2)_S$ , the duality transformations (\ref{dualtransf}) on the fields become
\begin{equation}
\begin{split}
& Y^0\rightarrow \Delta_S Y^0, \quad Y^1\rightarrow aY^1+bY^0\\
&Y^2 \rightarrow \Delta_S Y^2-\frac{2c}{Y^0}\frac{\partial \Omega}{\partial U},\\& Y^3 \rightarrow \Delta_S Y^3-\frac{2c}{Y^0}\frac{\partial \Omega}{\partial T}.
\end{split}
\end{equation}
In terms of special coordinates, these duality transformations become
\begin{eqnarray}\label{duality}
Y^0\rightarrow \Delta_S Y^0 , \qquad S\rightarrow \frac{aS-ib}{icS+d},\\\nonumber
T\rightarrow T +\frac{2ic}{\Delta_S(Y^0)^2}\frac{\partial\Omega}{\partial U} , \quad U\rightarrow U +\frac{2ic}{\Delta_S(Y^0)^2}\frac{\partial\Omega}{\partial T}.
\end{eqnarray}
Here
\begin{equation}
\Delta_S=d+icS
\end{equation}
One can similarly obtain the T(U)-duality transformation by interchanging the labels $1\leftrightarrow 2 (1\leftrightarrow 3)$.\\
Now requiring that the duality transformation is an invariance of the model and the transformation of $Y^I$ induce the transformation of $F_I$, the derivatives of $\Omega$ transform under $\Gamma(2)_S$ as \cite{0808.2627}
\begin{equation}
\left(\frac{\partial \Omega}{\partial T}\right)^{\prime}_S=\frac{\partial \Omega}{\partial T}  \qquad \left(\frac{\partial \Omega}{\partial U}\right)^{\prime}_S=\frac{\partial \Omega}{\partial U}\\\nonumber
\end{equation}
\begin{equation}
\left(\frac{\partial \Omega}{\partial S}\right)^{\prime}_S -\Delta^2_S\frac{\partial \Omega}{\partial S}=\frac{\partial(\Delta^2_S)}{\partial S}\left[-\frac{1}{2}Y^0\frac{\partial \Omega}{\partial Y^0}-\frac{ic}{\Delta_S(Y^0)^2}\frac{\partial \Omega}{\partial T}\frac{\partial \Omega}{\partial U}\right]\\\nonumber
\end{equation}
\begin{equation}\label{omegatransf}
\left(Y^0\frac{\partial \Omega}{\partial Y^0}\right)^{\prime}_S=Y^0\frac{\partial \Omega}{\partial Y^0}+\frac{4ic}{\Delta_S(Y^0)^2}\frac{\partial \Omega}{\partial T}\frac{\partial \Omega}{\partial U}.
\end{equation}
Similarly one can obtain the T(U)-duality transformation by replacing $S\leftrightarrow T (S\leftrightarrow U)$ in (\ref{omegatransf}).\\ 
Although for our analysis, we do not need explicit expression for $\Omega$,  we will just mention the results for $\Omega^{(1)}$ and $\Omega^{(2)}$ for STU-model. The result for $\Omega^{(1)}$ was derived in \cite{0808.2627}. One can follow the procedure of \cite{0808.2627} to get $\Omega^{(2)}$.
\begin{equation}
\begin{split}
&\Omega^{(1)}=\frac{1}{256\pi}\left[4ln[\vartheta_2(iS)\vartheta_2(iT)\vartheta_2(iU)]+c.c\right]+\frac{1}{126\pi}\left[ln[(S+\bar S)(T+\bar T)(U+\bar U)]\right]\\
&\Omega^{(2)}=\frac{2}{(Y^0)^2}\left[\frac{1}{S+\bar S}\frac{\partial \Omega^{(1)}}{\partial T}\frac{\partial \Omega^{(1)}}{\partial U}+\frac{1}{T+\bar T}\frac{\partial \Omega^{(1)}}{\partial U}\frac{\partial \Omega^{(1)}}{\partial S}+\frac{1}{U+\bar U}\frac{\partial \Omega^{(1)}}{\partial S}\frac{\partial \Omega^{(1)}}{\partial T}\right]+c.c
\end{split}
\end{equation}
where
\begin{equation}
\vartheta_2(iS)=\frac{2\eta^2(2iS)}{\eta(iS)}
\end{equation}
Here $\eta(iS)$ is the Dedekind function and $\eta^{24}(iS)$ is a modular form of weight 12 under SL(2;$\mathbb{Z}$). The expressions for $\Omega^{(1)}$ and $\Omega^{(2)}$ are determined upto $\Upsilon$ dependent S-, T-, U-duality invariant expressions.\\
Now we define the following $new$ variables
\begin{equation}\label{news}
S^{\prime}=S+\sum_{n=1}^{\infty}\Upsilon^n s_n(Y^{0},S,T,U,\bar{Y^0},\bar{S},\bar{T},\bar{U},\Omega)\\\nonumber
\end{equation}
\begin{equation}
T^{\prime}=T+\sum_{n=1}^{\infty}\Upsilon^n t_n(Y^{0},S,T,U,\bar{Y^0},\bar{S},\bar{T},\bar{U},\Omega)\\\nonumber
\end{equation}
\begin{equation}
U^{\prime}=U+\sum_{n=1}^{\infty}\Upsilon^n u_n(Y^{0},S,T,U,\bar{Y^0},\bar{S},\bar{T},\bar{U},\Omega)
\end{equation}
\begin{equation}
{Y^0}^{\prime}=Y^0+\sum_{n=1}^{\infty}\Upsilon^n k_n(Y^{0},S,T,U,\bar{Y^0},\bar{S},\bar{T},\bar{U},\Omega)\\\nonumber
\end{equation}
The above expansions are in positive powers of $\Upsilon$. The functions $\{s_n,t_n,u_n\}$ and $\{k_n\}$ are chosen such that under duality transformation (\ref{duality}) and (\ref{omegatransf}) the primed variables transform as a tree level variables i.e under $\Gamma(2)_S$ the primed variables should transform as 
\begin{equation}\label{dualityprime}
\begin{split}
&S^{\prime}\rightarrow \frac{aS^{\prime}-ib}{icS^{\prime}+d}, \qquad T^{\prime}\rightarrow T^{\prime}\\
&U^{\prime}\rightarrow U^{\prime}, \qquad {Y^0}^{\prime}\rightarrow (d+icS^{\prime}){Y^0}^{\prime}.
\end{split}
\end{equation}
Requiring this will give transformations of $s_n,t_n,u_n$ and $k_n$ from which one can determine $s_n$'s perturbatively in powers of $\Upsilon$.\\
\section{Determination of Field Redefinition}
In this section we will determine the form of $\{s_n,t_n,u_n\}$ and $\{k_n\}$ perturbatively. One of the advantages of considering STU-model is that if one obtains the expression of $s_n$, the expressions for $t_n$ and $u_n$ follow from triality. Furthermore, the expression for $k_n$ should be triality invariant. We will first determine the transformation of these functions and from there we will guess their form. However we found that from these transformations, one can not determine the form of these functions $uniquely$. There exist more than one such functions which give the same duality transformation (\ref{dualityprime}). This suggests that there are more than one such primed variables which under duality transformation do not mix with their complex conjugates. Here we mention the results for functions $\{s_n\}$ and $\{k_n\}$ only. The results for the other functions follow from triality.\\
\subsection{Determination of First Order Function}
Till first order in $\Upsilon$, the field redefinition are
\begin{equation}
\begin{split}
&S^{\prime}=S+\Upsilon s_1,\\
&{Y^0}^{\prime}=Y^0+\Upsilon k_1.
\end{split}
\end{equation}
Requiring that under old S-duality transformation (\ref{duality}), S$^{\prime}$ and ${Y^0}^{\prime}$ transform as (\ref{dualityprime}) gives the following duality transformations for $s_1$ and $k_1$
\begin{equation}
s_1\rightarrow \frac{s_1}{\Delta^2_S},\\\nonumber
\end{equation}
\begin{equation}\label{transf1}
k_1\rightarrow \Delta_S k_1 +ics_1 Y^0.
\end{equation}
Similar analysis under $T$ and $U$-duality gives the following results.\\
Under $T$-duality
\begin{equation}
s_1\rightarrow s_1-\frac{2ic}{(Y^0)^2\Delta_T}\left(\frac{\partial \Omega^{(1)}}{\partial U}\right).\\\nonumber
\end{equation}
Under $U$-duality 
\begin{equation}
s_1\rightarrow s_1-\frac{2ic}{(Y^0)^2\Delta_U}\left(\frac{\partial \Omega^{(1)}}{\partial T}\right).
\end{equation}
The transformation rule for $t_1$ $(u_1)$ will follow from the eqn.(\ref{transf1}) by interchanging $S \leftrightarrow T$ $(S\leftrightarrow U)$ and $\Gamma(2)_S \leftrightarrow \Gamma(2)_T$ $(\Gamma(2)_S \leftrightarrow \Gamma(2)_U)$.\\
Following are the transformation rule for $k_1$ under $T$ and $U$-duality. \\

Under $T$-duality
\begin{equation}
k_1\rightarrow \Delta_T k_1 +ict_1 Y^0.\\\nonumber
\end{equation}

Under $U$-duality
\begin{equation}\label{transk1}
k_1\rightarrow \Delta_U k_1 +icu_1 Y^0.
\end{equation}
From the above equation (\ref{transf1})- (\ref{transk1}) and (\ref{omegatransf}) one can guess the follwing expressions for $s_1$ and $k_1$\\
\begin{equation}\label{s_1}
s_1=\frac{2}{(Y^0)^2}\left[\frac{1}{T+\bar T}\left(\frac{\partial \Omega^{(1)}}{\partial U}\right)+\frac{1}{U+\bar U}\left(\frac{\partial \Omega^{(1)}}{\partial T}\right)\right],
\end{equation}
\begin{equation}
k_1=-\frac{Y^0}{2}\left[\frac{s_1}{S+\bar S}+\frac{t_1}{T+\bar T}+\frac{u_1}{U+\bar U}\right].
\end{equation}
The expression of $k_1$ is invariant under triality as required.\\
Similar expressions for $t_1$ and $u_1$ follows from triality. The expressions are
\begin{equation}
t_1=\frac{2}{(Y^0)^2}\left[\frac{1}{S+\bar S}\left(\frac{\partial \Omega^{(1)}}{\partial U}\right)+\frac{1}{U+\bar U}\left(\frac{\partial \Omega^{(1)}}{\partial S}\right)\right],
\end{equation}
\begin{equation}
u_1=\frac{2}{(Y^0)^2}\left[\frac{1}{T+\bar T}\left(\frac{\partial \Omega^{(1)}}{\partial S}\right)+\frac{1}{S+\bar S}\left(\frac{\partial \Omega^{(1)}}{\partial T}\right)\right].
\end{equation}
As we have mentioned earlier, there is an ambiguity in the obtained expression for $s_1$ (similarly one can find ambiguity in $t_1$, $u_1$ and $k_1$). One finds that one can add terms in $s_1$ keeping it's transformation rule intact. One such term is
\begin{equation}
f=\frac{1}{(Y^0)^2}\frac{\partial \Omega^{(1)}}{\partial U}\frac{\partial \Omega^{(1)}}{\partial T}.
\end{equation}
Under S-duality,
\begin{equation}
\frac{\partial \Omega^{(1)}}{\partial T}\rightarrow \frac{\partial \Omega^{(1)}}{\partial T}, \quad \frac{\partial \Omega^{(1)}}{\partial U}\rightarrow \frac{\partial \Omega^{(1)}}{\partial U} ,\quad Y^0\rightarrow \Delta_SY^0.
\end{equation}
Hence
\begin{equation}
f\rightarrow \frac{f}{\Delta_S^2}.
\end{equation}
Under T-duality,
\begin{equation}
\frac{\partial \Omega^{(1)}}{\partial T}\rightarrow\Delta_T^2 \frac{\partial \Omega^{(1)}}{\partial T} ,\quad \frac{\partial \Omega^{(1)}}{\partial U}\rightarrow \frac{\partial \Omega^{(1)}}{\partial U}, \quad Y^0\rightarrow \Delta_TY^0.
\end{equation}
Hence
\begin{equation}
f\rightarrow f.
\end{equation}
Similarly $f$ is invariant under U-duality also.\\
Hence one can add $f$ in $s_1$ and still we will have correct transformation for $s_1$. One can find the same ambiguity in other functions also. This ambiguity just reflects the fact that one can have more than one duality covariant complex variables which under duality transformation do not mix with their complex conjugates.
\subsection{Determination of Second Order Function}
In this section we repeat our calculations for second order function. We will do the analysis only for $s_2$ and $k_2$. One can find $t_2$ and $u_2$ by interchanging the fields.\\
We first need to find the duality transformation for $s_2$. Repeating the same procedure as we did for $s_1$, we get the following result\\
Under $S$-duality
\begin{equation}
\begin{split}
s_2\rightarrow &\frac{s_2}{\Delta^2_S}-\frac{ic}{\Delta^3_S}s_1^2-\frac{4ic}{\Delta^3_S(Y^0)^4}\frac{\partial \Omega^{(1)}}{\partial U}\left[-\frac{1}{(T+\bar T)^2}\frac{\partial \Omega^{(1)}}{\partial U}+\frac{1}{U+\bar U}\frac{\partial^2 \Omega^{(1)}}{\partial T^2}\right]\\& -\frac{4ic}{\Delta^3_S(Y^0)^4}\frac{\partial \Omega^{(1)}}{\partial T}\left[-\frac{1}{(U+\bar U)^2}\frac{\partial \Omega^{(1)}}{\partial T}+\frac{1}{T+\bar T}\frac{\partial^2 \Omega^{(1)}}{\partial U^2}\right]\\&+\frac{4ic}{\bar \Delta_S\Delta^2_S(Y^0\bar Y^0)^2} \frac{\partial \Omega^{(1)}}{\partial \bar U}\left[-\frac{1}{(T+\bar T)^2}\frac{\partial \Omega^{(1)}}{\partial U}+\frac{1}{U+\bar U}\frac{\partial^2 \Omega^{(1)}}{\partial T\partial {\bar T}}\right]\\&+\frac{4ic}{\bar \Delta_S\Delta^2_S(Y^0\bar Y^0)^2} \frac{\partial \Omega^{(1)}}{\partial \bar T}\left[-\frac{1}{(U+\bar U)^2}\frac{\partial \Omega^{(1)}}{\partial T}+\frac{1}{T+\bar T}\frac{\partial^2 \Omega^{(1)}}{\partial U\partial {\bar U}}\right].
\end{split}
\end{equation}
Under $T$-duality
\begin{equation}
\begin{split}
s_2\rightarrow & s_2 -\frac{2ic}{\Delta_T(Y^0)^2}\frac{\partial \Omega^{(2)}}{\partial U}-\frac{4ic}{\Delta_T(Y^0)^4}\frac{\partial \Omega^{(1)}}{\partial S}\left[\frac{1}{T+\bar T}\frac{\partial^2\Omega^{(1)}}{\partial U^2}-\frac{1}{(U+\bar U)^2}\frac{\partial \Omega^{(1)}}{\partial T}\right]\\&+\frac{4ic}{\bar \Delta_T(Y^0\bar Y^0)^2}\frac{\partial \Omega^{(1)}}{\partial \bar S}\left[\frac{1}{T+\bar T}\frac{\partial^2\Omega^{(1)}}{\partial U \partial \bar U}-\frac{1}{(U+\bar U)^2}\frac{\partial \Omega^{(1)}}{\partial T}\right]\\&-\frac{4c^2}{\Delta_T^2(Y^0)^4}\frac{\partial \Omega^{(1)}}{\partial S}\frac{\partial^2 \Omega^{(1)}}{\partial U^2}+\frac{4c^2}{\Delta_T \bar \Delta_T(Y^0\bar Y^0)^2}\frac{\partial \Omega^{(1)}}{\partial \bar S}\frac{\partial^2 \Omega^{(1)}}{\partial U \partial \bar U}.
\end{split}
\end{equation}
We have similar expression under $U$-duality with $T\leftrightarrow U$ in the above expression for T-duality.\\
The expression for $s_2$-consistent with above transformation is
\begin{equation}
\begin{split}
s_2=&\frac{2}{(Y^0)^2}\left[\frac{1}{U+\bar U}\frac{\partial \Omega^{(2)}}{\partial T}+\frac{1}{T+\bar T}\frac{\partial \Omega^{(2)}}{\partial U}\right]-\frac{2}{Y^0}\frac{1}{(T+\bar T)(U+\bar U)}\frac{\partial \Omega^{(2)}}{\partial Y^0}\\&-\frac{2s_1}{(T+\bar T)(U+\bar U)}\left[\frac{1}{(Y^0)^2}\frac{\partial \Omega^{(1)}}{\partial S}+\frac{1}{(\bar Y^0)^2}\frac{\partial \Omega^{(1)}}{\partial \bar S}\right]\\& -\frac{4}{(S+\bar S)(Y^0\bar Y^0)^2}\left[\frac{1}{(T+\bar T)^2}\frac{\partial \Omega^{(1)}}{\partial \bar U}\frac{\partial \Omega^{(1)}}{\partial U}+\frac{1}{(U+\bar U)^2}\frac{\partial \Omega^{(1)}}{\partial \bar T}\frac{\partial \Omega^{(1)}}{\partial T}\right].
\end{split}
\end{equation}
The expressions for $t_2$ and $u_2$ will follow from triality symmetry.\\
Similarly one can obtain the expression for $k_2$. We will not write here the transformation of $k_2$,  but just mention the result for $k_2$
\begin{equation}
\begin{split}
k_2=&-\frac{Y^0}{2}\left[\frac{s_2}{S+\bar S}+\frac{t_2}{T+\bar T}+\frac{u_2}{U+\bar U}\right]+\frac{1}{(S+\bar S)(T+\bar T)(U+\bar U)}\frac{\partial \Omega^{(2)}}{\partial Y^0}\\&+\frac{Y^0}{2}\left[\frac{s_1^2}{(S+\bar S)^2}+\frac{t_1^2}{(T+\bar T)^2}+\frac{u_1^2}{(U+\bar U)^2}\right]\\&+\frac{1}{Y^0(S+\bar S)(T+\bar T)(U+\bar U)}\left[(s_1+\bar s_1)\frac{\partial \Omega^{(1)}}{\partial S}+(t_1+\bar t_1)\frac{\partial \Omega^{(1)}}{\partial T}+(u_1+\bar u_1)\frac{\partial \Omega^{(1)}}{\partial U}\right].
\end{split}
\end{equation}
The expression for $k_2$ is also triality invariant.
\section{Free Energy}
In this section we express the free energy as a function of the new variables. The expression for free energy (\ref{modfrenrg}) in terms of old variables upto first order in $\Upsilon$ is
\begin{equation}
\begin{split}
\mathscr F=&-\vert Y^0\vert^2(S+\bar S)(T+\bar T)(U+\bar U)+4\Upsilon\Omega^{(1)}\\&-2\Upsilon\left\{\frac{\bar Y^0}{Y^0}\left[(S+\bar S)\frac{\partial \Omega^{(1)}}{\partial S}+(T+\bar T)\frac{\partial \Omega^{(1)}}{\partial T}+(U+\bar U)\frac{\partial \Omega^{(1)}}{\partial U}\right]+h.c\right\}.
\end{split}
\end{equation}

The above expression written in terms of new variables is
\begin{equation}
\mathscr F=-{\vert {Y^0}^{\prime}\vert}^2(S^{\prime}+\bar S^{\prime})(T^{\prime}+\bar T^{\prime})(U^{\prime}+\bar U^{\prime})+4\Upsilon{\Omega^{(1)}}^{\prime}+O(\Upsilon^2).
\end{equation}
where
\begin{equation}
\Omega^{(1)\prime}=\frac{1}{256\pi}\left[4ln[\vartheta_2(iS^{\prime})\vartheta_2(iT^{\prime})\vartheta_2(iU^{\prime})]+c.c\right]+\frac{1}{126\pi}\left[ln[(S^{\prime}+\bar S^{\prime})(T^{\prime}+\bar T^{\prime})(U^{\prime}+\bar U^{\prime})]\right].
\end{equation}
One can check that the leading and the $\Upsilon$ dependent terms are separately invariant under the duality transformation(\ref{dualityprime}).
\section{Discussion}
It has been demonstrated in \cite{0808.2627,9906094,0601108} that the nonholomorphic corrections are necessary in order to obtain duality invariant expressions for the free energy and the entropy function. But at the same time, our present method \cite{0808.2627} of incorporating such corrections is manifestly in conflict with the special geometry properties of the moduli space. The moduli fields mix with their complex conjugates under duality transformation and moreover these transformations involve powers of $\Upsilon$. In this note we have shown that it is possible to find a set of new variables  perturbatively in powers of $\Upsilon$ which have the usual tree-level duality transformation (\ref{dualityprime}). We have demonstrated their existence upto $\Upsilon^2$ order in the STU-model, but it seems likely that one can find such variables in any model and to any order in $\Upsilon$. Furthermore, these primed varibles are more natural from the point of view of string theory and topological string. The reason is that the partition function for the STU- model propsed in \cite{0711.1971} involves the moduli which transform in the usual manner under the S-T-U duality transformation (\ref{dualityprime}). The primed variables are also more natural in the context of topological string because the integrability of the holomorphic anamoly equations depends on the special geometry properties of the moduli space\cite{9309140}.
Furthermore in these new varibles the expression of the free energy ,at each order in $\Upsilon$, is invariant under tree level duality transformation (\ref{dualityprime}) and hence is the natural candidate to compare with the topological string result for the free energy \cite{0702187}.\\
{\bf Acknowledgement:} We would like to thank Arjun Bagchi, Justin David, Dileep Jatkar and Ashoke Sen for various useful discussion. We would specially like to thank Ashoke Sen for taking intrest in our work from beginning and also useful comments on the first draft.

\end{document}